\documentclass[12pt,aps,pra,superscriptaddress,preprint,groupedaddress,showpacs]{revtex4}
\usepackage{graphicx}
\usepackage{amsmath,amssymb,amsfonts}
\usepackage{natbib}

\newcommand{\be}{\begin{equation}}
\newcommand{\ee}{\end{equation}}
\newcommand{\bd}{\begin{displaymath}}
\newcommand{\ed}{\end{displaymath}}
\newcommand{\bq}{\begin{eqnarray}}
\newcommand{\eq}{\end{eqnarray}}

\begin{document}

\title{Contextuality, decoherence and quantum trajectories}

\author{A. S. Sanz}
\email{asanz@imaff.cfmac.csic.es}

\affiliation{Instituto de F\'{\i}sica Fundamental,\\
Consejo Superior de Investigaciones Cient\'{\i}ficas,\\
Serrano 123, 28006 Madrid, Spain}

\author{F. Borondo}
\email{f.borondo@uam.es}

\affiliation{Instituto Mixto de Ciencias Matem\'aticas
CSIC--UAM--UC3M--UCM, and
Departamento de Qu\'\i mica, C--IX, \\
Universidad Aut\'onoma de Madrid,
Cantoblanco -- 28049 Madrid, Spain}

\date{\today}

\begin{abstract}
Here we analyze the relationship between quantum contextuality and
decoherence in interference experiments with matter particles by
means of a simple reduced quantum-trajectory model, which attempts
to simulate the behavior of the projections of multi-dimensional,
system-plus-environment Bohmian trajectories onto the subspace of
the reduced system.
This model allows us to understand the crossing of the subsystem
trajectories as a combined effect of interference quenching and erasure
of ``which-way'' information, which can be of utility to interpret
decoherence effects in many-dimensional systems where full Bohmian
treatments become prohibitive computationally.
\end{abstract}



\maketitle


\section{\label{sec1} Introduction}

Quantum physics is characterized by striking properties which puzzle
and challenge our understanding of the physical world.
One of them is {\it contextuality}, the unavoidable dependence of
the description of a system on the experimental (or contextual) setup,
which strongly determines the wave function describing the quantum
state of that system \cite{clementetorre}.
This strongly contrasts with the behavior displayed by classical
systems, whose dynamics is only governed by the eventual external
forces.
The interference of separate diffracted beams of matter particles
\cite{Davisson,Zeilinger1,tonomura,shimizu,Zeilinger2,urena1,urena2}
constitutes a nice scenario where this property becomes very apparent.
Consider, for instance, a two-slit diffraction experiment.
If no ``which-way'' information is demanded, the diffracted projectiles
will distribute beyond the slits at a time $t$ according to the
probability density
\be
 \rho_t ({\bf r}) = |\psi_{1,t}({\bf r}) + \psi_{2,t}({\bf r})|^2
   = \rho_{1,t}({\bf r}) + \rho_{2,t}({\bf r})
     + \rho_{{\rm int},t}({\bf r}) .
 \label{rel1}
\ee
Here, $\psi_{i,t}({\bf r})$ ($i=1,2$) is the wave function describing
the particle after {\it only} passing through slit $i$ and
$\rho_{i,t}({\bf r})=|\psi_{i,t}({\bf r})|^2$ is the probability
corresponding to finding that particle at ${\bf r}$.
The particular fringe--like features observed in the interference
pattern arise from the interference term $\rho_{{\rm int},t}({\bf r})$,
which is associated with the contextual or experimental information
``both slits are open simultaneously when the particle beam reaches
them''.
Otherwise, if we choose our context to be one slit open at a time,
particles will distribute according to the classical-like probability
distribution
\be
 \rho_t({\bf r}) = |\psi_{1,t}({\bf r})|^2 + |\psi_{2,t}({\bf r})|^2
   = \rho_{1,t}({\bf r}) + \rho_{2,t}({\bf r}) ,
 \label{rel2}
\ee
which obviously does not contain the interference term.
Thus, for the same physical system, two different contexts provide two
different answers ---note that in classical mechanics we would only
have the case described by Eq.~(\ref{rel2}), since the (conditional)
probability (density) to pass through one of the slits when the other
is also open is just zero.

In the literature it is common to associate the two contexts
described above with either the duality principle (with the
two slits open simultaneously, the system behaves as a wave;
with only one, as a corpuscle) or the uncertainty principle
(determining the particle position is the same as to determine
which slit the particle passed through, while letting it pass
through both slits open allows us to determine its momentum).
However, we can also find a relationship between the change of
context and the problem of decoherence.
As is well known, decoherence is nowadays the most widespread mechanism
used in the literature \cite{Giulini1} to explain the appearance of
classical-like behaviors in quantum systems due to their dynamical
interaction with an environment.
If system plus environment are initially described by a separable total
wave function, their coupling makes this wave function to become
{\it entangled} \cite{schro1,schro2} in time.
When we ``look'' at the reduced system dynamics, this entanglement
translates into a partial or even total loss of the system coherence.
In the case of the two slits, this process could be seen as a sort
of smooth transition from a context where both slits are open
simultaneously, to another in which only one slit is open at a
time, because of the gradual {\it screening} of the information which
establishes that both were open when the particle got diffracted.
In other words, as the entanglement becomes stronger, the effect is
similar to assume that the particle track becomes more localized in
time.

The transition from one context to the other due to decoherence can be
very well visualized in terms of quantum trajectories \cite{Bohm1,Bohm2},
since they offer a clear interpretational advantage by making
possible to follow the system dynamics and to (intuitively)
understand the underlying physics of the process at the same
level that classical trajectories do in classical mechanics.
It is well known that the standard version of quantum mechanics
(Copenhagen interpretation) can be reformulated in
an exact fashion in terms of a trajectory-based theory, namely Bohmian
mechanics \cite{Bohm1,Bohm2,Holland}, which has been widely used,
precisely, to describe quantum interference and diffraction
\cite{Philippidis1,Philippidis2,Sanz0,Sanz1,Sanz2} among many
other different applications.
However, and as it also happens in standard quantum mechanics,
Bohmian mechanics cannot be directly applied to decoherence
problems with a large number of degrees of freedom because any
calculation becomes numerically prohibitive \cite{wyatt-book}.
This inconvenience can be avoided, nevertheless, by following
the same philosophy as in standard quantum mechanics when
Markovian conditions apply.
In such cases, the effects of the environment on the system dynamics
appear through some dissipative terms ({\it dissipators}) in the system
equations of motion, as can be seen in different approaches based on
solving either master equations for the reduced density matrix
\cite{Breuer} or stochastic wave equations \cite{percival}.
This allows one to devise simple, phenomenological models to
understand the way how the coherence damping takes place
\cite{tan-walls,Savage1,Savage2,Sanz31,Sanz32,qureshi}.
Similarly, one can also devise simple quantum trajectory models
grounded on Bohmian mechanics, such as the reduced quantum trajectory
formalism \cite{SanzEPJD}, where the environment effects appear in the
system equations of motion through some dissipative terms and,
therefore, it is not necessary to deal with the dynamics of the
whole system-plus-environment set of degrees of freedom.
In this regard, we would like to mention that former analyses of
decoherence by means of quantum trajectories were carried out by Na
and Wyatt \cite{na1,na2}, where a coherent superposition was coupled
to a harmonic bath.

Finally, it is worth pointing out that the questions addressed
here, i.e.\ entanglement and decoherence, have recently received
a great deal of attention from the theoretical chemistry community.
This arises from the feasibility of using excited vibrational states
as a basis for quantum computing \cite{chemistry1a,chemistry1b,%
chemistry1c}.
Then problems such as entanglement dynamics \cite{chemistry2a,%
chemistry2b,chemistry2c,chemistry2d},
or the role of the different characteristics of the corresponding
potential energy surfaces \cite{chemistry3} have been addressed in
molecular states in H$_2$O, SO$_2$, O$_3$, and formaldehyde.

In this work we explore the properties of reduced quantum trajectory
formalisms in connection with the problems of contextuality and
decoherence.
In particular, we propose a simple trajectory model which includes
also a simple mechanism of information screening.
Although approximate, the quantum trajectories within this model are
able to reproduce satisfactorily the projection of the ``true''
Bohmian trajectories onto the subspace of the reduced system,
since it can be expected that, due to entanglement,
such projections violate the non-crossing property of
Bohmian mechanics \cite{jpa}.
To make this paper self-contained, prior to the description of our
model, we briefly analyze in Sec.~\ref{sec2} the problem of
decoherence in two-slit experiments from the standard quantum-mechanical
point of view.
Then, in Sec.~\ref{sec3} we describe the reduced quantum trajectory
model based on the screening mechanism mentioned above.
Numerical simulations based on this model are presented and discussed
in Sec.~\ref{sec4}.
A final discussion and the main conclusions derived from the present
work are summarized in Sec.~\ref{sec5}.


\section{\label{sec2} Decoherence in the double-slit experiment}

Consider a particle beam after getting diffracted by two slits.
In the absence of interactions with an environment, its time-evolution
can be described at any time by
\begin{equation}
 |\Psi^{(0)}_t\rangle
   = c_1 |\psi_{1,t}\rangle + c_2 |\psi_{2,t}\rangle ,
 \label{eq1}
\end{equation}
where $|c_1|^2 + |c_2|^2 = 1$.
In the coordinate representation, the element $({\bf r},{\bf r}')$ of
the associate density matrix, $\hat{\rho}_t^{(0)}=|\Psi^{(0)}_t\rangle
\langle\Psi^{(0)}_t|$, will read as
\begin{equation}
 \rho_t^{(0)} ({\bf r},{\bf r}') =
  \Psi^{(0)}_t ({\bf r}) \left[ \Psi^{(0)}_t ({\bf r}') \right]^* ,
 \label{eq2}
\end{equation}
and the diagonal terms giving the probability density as
\be
 \rho_t^{(0)} ({\bf r})
   = |c_1|^2 |\psi_{1,t}({\bf r})|^2 + |c_2|^2 |\psi_{2,t}({\bf r})|^2
     + 2 |c_1| |c_2| |\psi_{1,t}({\bf r})| |\psi_{2,t}({\bf r})|
        \cos \delta_t ({\bf r}) ,
 \label{eq3}
\ee
with $\Psi^{(0)}_t({\bf r}) = \langle {\bf r} |\Psi^{(0)}_t\rangle$.
In this last expression, $\delta_t$ is the space and time dependent
phase-shift between the two partial waves.

In order to include the effects of an environment over the system,
consider now the following simple model (though general enough to
study other interference process than two-slit experiments).
Under the presence of such an environment, Eq.~(\ref{eq1}) is no longer
valid to describe the system dynamics.
Assuming that system and environment are initially decoupled, the
(initial) total wave function accounting for both can be expressed as
a factorized product of the initial wave function describing each
subsystem,
\begin{equation}
 |\Psi \rangle = |\Psi^{(0)}\rangle \otimes |E_0\rangle ,
 \label{eq4}
\end{equation}
with $|\Psi^{(0)}\rangle$, as in Eq.~(\ref{eq1}).
If the environment acts differently on each branch of the diffracted
beam \cite{Giulini1,tan-walls,Sanz31,Sanz32}, at any subsequent time
we will find
\begin{equation}
 |\Psi_t \rangle = c_1 |\psi_{1,t}\rangle \otimes |E_{1,t}\rangle
  + c_2 |\psi_{2,t}\rangle \otimes |E_{2,t}\rangle ,
 \label{eq5}
\end{equation}
i.e.\ the total wave function has become {\it entangled}.
The reduced probability density associated with the system of interest
is now obtained from (\ref{eq5}) by tracing the full density matrix,
$\hat{\rho}_t = |\Psi_t\rangle \langle\Psi_t|$, over the environment
states, which leads to
\begin{equation}
 \hat{\tilde{{\rho}}}_t =
  \sum_{j=1,2} \langle E_{j,t} | \hat{\rho}_t | E_{j,t} \rangle .
 \label{eq6}
\end{equation}
Note that, in the coordinate representation and for an environment
constituted by $N$ particles, Eq.~(\ref{eq6}) becomes
\be
 \tilde{\rho}_t ({\bf r}, {\bf r}') = \int
  \langle {\bf r}, {\bf r}_1, {\bf r}_2, \ldots {\bf r}_N |
    \Psi_t \rangle
    \langle \Psi_t
  | {\bf r}', {\bf r}_1, {\bf r}_2, \ldots {\bf r}_N \rangle
  \ \! {\rm d}{\bf r}_1 {\rm d}{\bf r}_2 \cdots {\rm d}{\bf r}_N .
 \label{eq7}
\ee
Substituting (\ref{eq5}) into Eq.~(\ref{eq6}), rearranging terms
and then expressing the final result in the reduced coordinate
representation, we reach
\be
 \tilde{\rho}_t ({\bf r}, {\bf r}') =
 (1 + |\alpha_t|^2) \sum_{j=1,2}
  |c_j|^2 \psi_{j,t} ({\bf r}) \psi_{j,t}^* ({\bf r}')
  + 2 \alpha_t c_1 c_2^* \psi_{1,t} ({\bf r}) \psi_{2,t}^* ({\bf r}')
  + c.c. ,
 \label{eq8}
\ee
where $\alpha_t = \langle E_{2,t} | E_{1,t} \rangle$ and $c.c.$
indicates the complex conjugate of the second term on the r.h.s.
Finally, the (reduced) probability density resulting from
Eq.~(\ref{eq8}) is
\be
 \tilde{\rho}_t = (1 + |\alpha_t|^2)
   \left[ |c_1|^2 |\psi_{1,t}|^2 + |c_2|^2 |\psi_{2,t}|^2
    + 2 \Lambda_t |c_1| |c_2| |\psi_{1,t}| |\psi_{2,t}|
        \cos \delta'_t \right] ,
 \label{eq9}
\ee
with
\begin{equation}
 \Lambda_t = \frac{2 |\alpha_t|}{(1 + |\alpha_t|^2)}
 \label{eq10}
\end{equation}
being the {\it coherence degree} \cite{Sanz31,Sanz32}, which gives an
idea of the {\it fringe visibility} of the interference pattern.
For instance, if we consider $|\alpha_t|=e^{-t/\tau_c}$, we find
\begin{equation}
 \Lambda_t = {\rm sech} (t/\tau_c) ,
 \label{eq11}
\end{equation}
where $\tau_c$ is the {\it coherence time}, which is a function of
different physical parameters (the system mass, temperature, etc.).
In the literature one can find detailed models which allow one to get
an estimation of coherence times in different physical situations
\cite{Savage1,Savage2,qureshi}.
Nevertheless, as can be seen in Eq.~(\ref{eq9}), if $\tau_c\to\infty$,
the decoherence process is very slow and the interference pattern is
always well defined.
On the contrary, for $\tau_c$ finite we observe an asymptotic decay of
the interference pattern to a classical-like one described by
\begin{equation}
 \tilde{\rho}_t = |c_1|^2 |\psi_{1,t}|^2 + |c_2|^2 |\psi_{2,t}|^2 .
 \label{eq9bis}
\end{equation}
That is, we observe a smooth transition from the context where both
slits were open simultaneously to another one where only one slit is
open at a time, not precisely because this was the case, but because
the environment makes the information of simultaneity to become
screened.


\section{\label{sec3} A simple reduced quantum trajectory model for
loss of ``which-way'' information}

In order to visualize and describe the action of the environment over
the system in terms of trajectories, it is convenient to express the
probability current density ${\bf J}_t$ in terms of the system density
matrix,
\begin{equation}
 {\bf J}_t = \frac{\hbar}{m}
  \ {\rm Im} [ \nabla_{\bf r} {\rho}_t^{(0)} ({\bf r},{\bf r}')]
  \Big\arrowvert_{{\bf r}' = {\bf r}} ,
 \label{subeq12}
\end{equation}
instead of starting from the standard Bohmian derivation in terms of
$\Psi^{(0)}_t$ \cite{jpa}.
Thus, under the presence of the environment, Eq.~(\ref{subeq12})
becomes
\begin{equation}
 \tilde{\bf J}_t \equiv \frac{\hbar}{m}
  \ {\rm Im} [ \nabla_{\bf r} \tilde{\rho}_t ({\bf r},{\bf r}')]
  \Big\arrowvert_{{\bf r}' = {\bf r}} ,
 \label{eq12}
\end{equation}
which satisfies the (reduced) continuity equation
\begin{equation}
 \frac{\partial \tilde{\rho}_t}{\partial t}
   + \nabla \tilde{\bf J}_t = 0 .
 \label{eq13}
\end{equation}
After Eqs.~(\ref{eq12}) and (\ref{eq13}), one can think of the reduced
probability current density as a transport effect of the reduced
probability density through a (reduced) velocity field,
$\tilde{\bf v}$, according to the relation
\begin{equation}
 \tilde{\bf J}_t = \tilde{\rho}_t \tilde{\bf v} .
 \label{vfield}
\end{equation}
This field is the reduced analog of the Bohmian velocity field
and we can obtain reduced quantum trajectories from it by defining
\cite{SanzEPJD}, in analogy to Bohmian mechanics, the equation of
motion
\begin{equation}
 \tilde{\bf v} = \dot{\bf r}_t = \frac{\hbar}{m}
  \frac{{\rm Im} [ \nabla_{\bf r} \tilde{\rho}_t ({\bf r},{\bf r}')]}
       {{\rm Re} [ \tilde{\rho}_t ({\bf r},{\bf r}')]}
   \Bigg\arrowvert_{{\bf r}' = {\bf r}} .
 \label{eq14}
\end{equation}

Due to the continuity equation (\ref{eq13}) and definition
(\ref{vfield}), the dynamics described by Eq.~(\ref{eq14}) leads to
the correct intensity pattern when the statistics of a large particle
ensemble is considered \cite{SanzEPJD}, as also happens in standard
Bohmian mechanics.
However, despite the insight on decoherence processes provided by this
equation as well as its computational advantages, a close analysis
shows us that it still keeps the non-crossing property of Bohmian
mechanics.
If we consider the limit $t \gg \tau_c$ ($\alpha_t \to 0$),
Eq.~(\ref{eq14}) becomes
\begin{equation}
 \dot{\bf r}_t = \frac{|c_1|^2 \rho_{1,t} \dot{\bf r}_{1,t} +
   |c_2|^2 \rho_{2,t} \dot{\bf r}_{2,t}}{\rho_{{\rm cl},t}} ,
 \label{eq18}
\end{equation}
where $\dot{\bf r}_{j,t}$ and $\rho_{j,t}$ are, respectively, the
velocity field and probability density associated with the wave
$|\psi_{j,i}\rangle$, and
\begin{equation}
 \rho_{{\rm cl},t} \equiv |c_1|^2 \rho_{1,t} + |c_2|^2 \rho_{2,t} .
 \label{rhodec}
\end{equation}
Note that, in this limit, both $\rho_{{\rm cl},t}$ and the probability
current density,
\begin{equation}
 {\bf J}_{{\rm cl},t} \equiv \rho_{{\rm cl},t} \ \dot{\bf r}_t =
  |c_1|^2 \rho_{1,t} \ \dot{\bf r}_{1,t} +
  |c_2|^2 \rho_{2,t} \ \dot{\bf r}_{2,t} ,
 \label{eq19}
\end{equation}
are properly defined.
However, Eq.~(\ref{eq18}) still contains information on both slits,
which leads to the non-crossing of the reduced trajectories, while the
expected behavior would be just a crossing as will happen with the
projection of the true Bohmian trajectories on the subspace of the
reduced system \cite{marchildon}.
In order to observe a full transition towards a classical-like regime
within the theoretical framework of the model here described, it is
therefore necessary the gradual screening of such an information.

In order to observe such a behavior, we can further proceed with our
model as follows.
If, from a trajectory point of view, a particle does not pass through
one slit, we will call this the empty slit.
Thus, in analogy to $\alpha_t$, assume that the influence of the
empty slit on the particle motion decreases exponentially due to
the increasing entanglement with the environment.
At the same time, the loss of information about the empty slit
strengthes the information about the traversed one.
In other words, the weight or influence of the empty slit decreases
with time in the corresponding trajectory evolution, while that of the
traversed slit increases.
Consider that the decay of information goes exponentially.
That is, if the particle passes through, say, slit 1, we assume that
the coefficient associated with $|\psi_{2,t}\rangle$ is given by
$c'_2 = c_2 \ \! {\rm e}^{-t/\tau_s}$, where $\tau_s$ is the
{\it screening time}, in analogy to the coherence time $\tau_c$.
Here, note that $\tau_s$ gives a timescale related to how fast the
information provided by the empty slit is screened or decoupled from
the particle motion due to the environment.
On the other hand, for the coefficient of $|\psi_{1,t}\rangle$ we
choose
\be
 c'_1 = c_1 \sqrt{(1 - |c_2|^2 {\rm e}^{-2t/\tau_s})/|c_1|^2} ,
 \label{coeff1}
\ee
indicating the increasing role of the traversed slit in the quantum
motion.
Now, the evolution of the system is then described by Eq.~(\ref{eq14}),
but replacing the $c_i$ by their respective time-dependent
counterparts, $c'_i$.

Since there are two characteristic times, $\tau_s$ and $\tau_c$, we can
define the ratio $\eta \equiv \tau_s/\tau_c$.
Thus, if $\eta \ll 1$, the screening of the empty-slit information
takes place much faster than the process that leads to the quenching
or damping of the interference fringes.
In this case, if the screened slit is, say, 2, Eq.~(\ref{eq14})
reduces to
\begin{equation}
 \dot{\bf r}_t = \dot{\bf r}_{1,t}
 \label{eq20}
\end{equation}
and the trajectories will evolve like if there was no other slit at
all, i.e.\ like in a context where there is only one slit open at a
time.
This means that particles are allowed to cross the symmetry axis of
the experiment because, at a given time, the momentum can have two
different values on the same space point, as in classical mechanics.
In next section we illustrate this processes by means of some numerical
simulations.


\section{\label{sec4} Numerical simulations}

\begin{figure}
 \includegraphics[width=12cm]{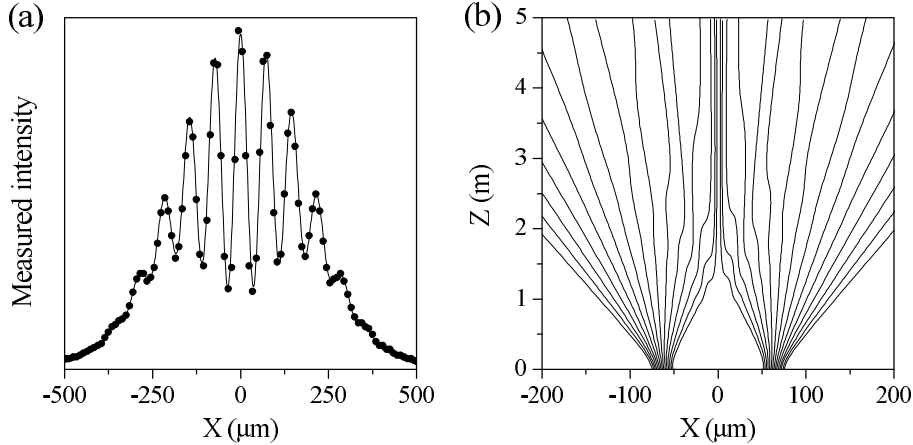}
 \caption{\label{fig:1} a) Intensity obtained from the statistics
 of reduced quantum trajectories (full circles) and standard quantum
 mechanics (solid line) with $\tau_c = 2.26\times10^{-2}$~s.
 b) Sample of reduced quantum trajectories illustrating the dynamics
 associated with the results shown in part (a).}
\end{figure}

The parameters (dimensions of the two-slit assembly, particle masses
and wavelengths) utilized in the simulations that we will present
here refer to the two-slit experiment with cold neutrons carried out
by Zeilinger {\it et al.}\ \cite{Zeilinger1}, which we will utilize
as a working model.
Thus, in Fig.~\ref{fig:1}(a), we show the excellent agreement between
the statistics over reduced quantum trajectories (full circles) and the
corresponding standard quantum mechanical calculations (solid line),
in accordance to the continuity equation (\ref{eq13}) and definition
(\ref{vfield}) for the reduced field.
A sample of reduced trajectories illustrating the dynamics associated
with the results of Fig.~\ref{fig:1}(a) is displayed in part (b) of the
same figure.
Here, we see that as the outgoing neutron beams start to interfere
(at a distance of $\sim$1~m from the two slits), some trajectories
(mainly those closer to the symmetry axis of the experiment) start
to show a conspicuous change of direction, which is typical in the
Bohmian description of interference processes
\cite{Philippidis1,Sanz0,Sanz1}.
However, due to the interference quenching, the extent of this behavior
is reduced in both space and time: in space because the interference
effects are stronger for the innermost trajectories [which give rise
to the central peaks in Fig.~\ref{fig:1}(a)] and in time because the
time-of-flight of the neutrons ($\tau_f = 2.33\times10^{-2}$~s) is
slightly larger than $\tau_c$.

\begin{figure}
 \includegraphics[width=12cm]{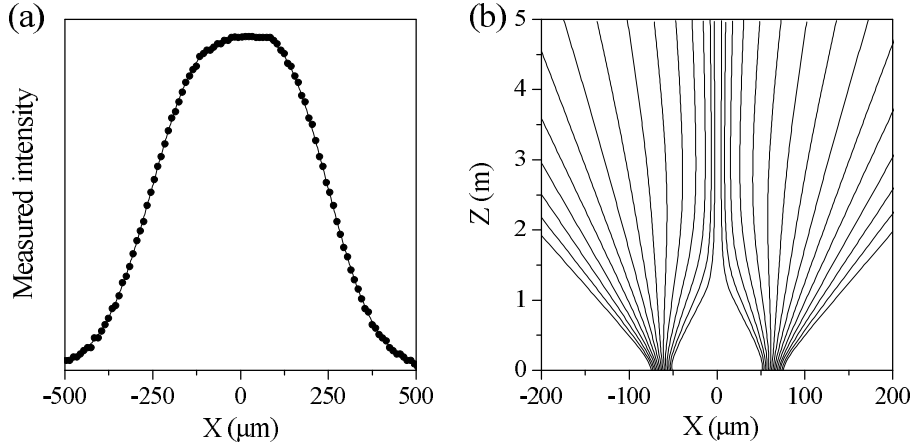}
 \caption{\label{fig:2} a) Intensity obtained from reduced quantum
 trajectory (full circles) and standard quantum mechanics (solid line)
 calculations for a double-slit experiment with cold neutrons and
 $\tau_c = 0$ (no coherence).
 b) Sample of reduced quantum trajectories illustrating the dynamics
 associated with the results shown in part (a).}
\end{figure}

In Fig.~\ref{fig:2} we show the results for the extreme case of maximal
decoherence, that takes place for $\tau_c = 0$.
Similarly to what we did in Fig.~\ref{fig:1}, the statistical results
obtained by means of quantum trajectories and standard quantum mechanics (a),
as well as a sample of representative trajectories (b) are plotted.
Notice that despite there is no coherence (in the sense that the
interference terms of the reduced density matrix have been damped out),
trajectories do not cross the symmetry axis between the two slits
because they obey Eq.~(\ref{eq18}), which contains information about
the two slits open simultaneously.
The absence of interference only prevents the particles from undergoing
the typical ``wiggling'' motion leading to the different diffraction
peaks \cite{Sanz1}.

\begin{figure*}
 \includegraphics[width=16cm]{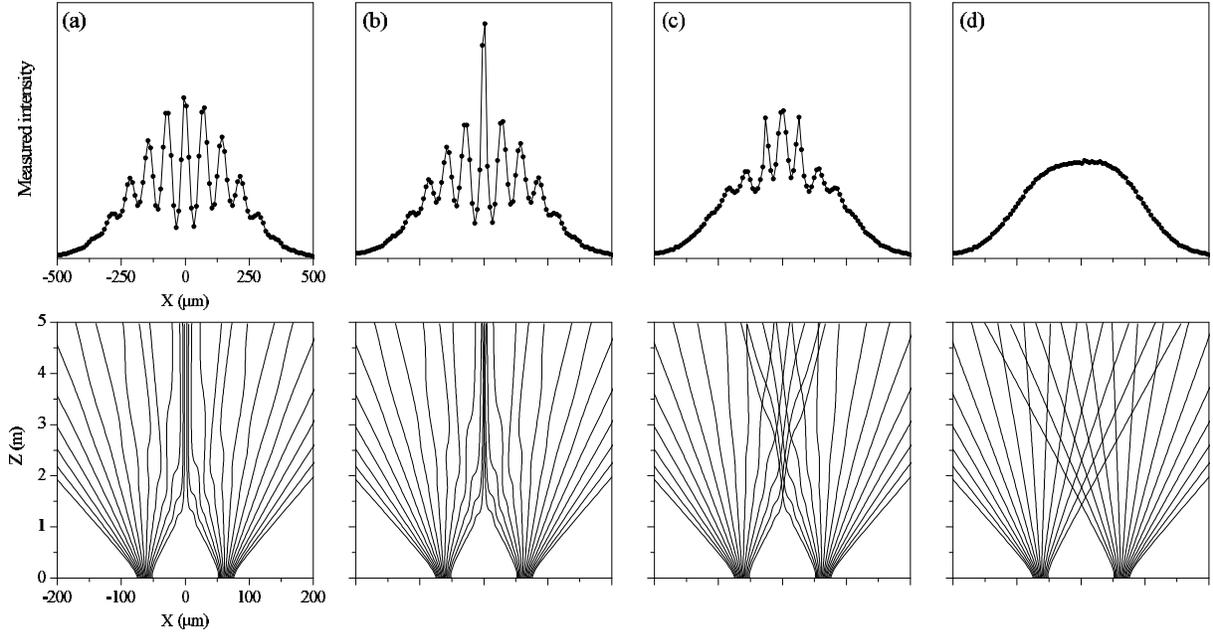}
 \caption{\label{fig:3} Top: Intensity obtained after counting
 of the corresponding pseudo-reduced quantum trajectories for:
 a) $\eta = \infty$, b) $\eta = 10$, c) $\eta = 1$, and
 d) $\eta = 0.1$.
 The results from the counting (full circles) have been joined by means
 of B-splines (solid line) in order to facilitate their understanding.
 Bottom: Samples of trajectories illustrating the dynamics of the
 results shown in the upper part.}
\end{figure*}

In order to illustrate how the additional mechanism leading to the
screening of the information provided by the empty slit works, we show
in Fig.~\ref{fig:3} results for several values of the parameter $\eta$
once the $c_i$ coefficients have been replaced by the $c'_1$ ones in
Eq.~(\ref{eq14}).
The statistics of reduced-screened quantum trajectories are given in
the top row and samples of quantum trajectories illustrating the
corresponding dynamics are displayed in the bottom one.
As can be noticed in Fig.~\ref{fig:3}(a), for small values of $\eta$
the screening is very slow screening and the non-crossing dominates
the system dynamics.
However, as we move to higher values of $\eta$ (from left to right in
the figure), the screening starts to increase in importance over the
interference quenching and trajectories start to cross the axis
between the two slits.
In the extreme case, where the screening is very strong, the
trajectories display a classical-like behavior because the information
about the two slits open simultaneously initially is lost very rapidly.

The transition from one regime to the other one is also interesting
both from the dynamical and the statistical points of view.
As $\eta$ increases an order of magnitude, a set of trajectories
densely concentrates along the symmetry axis, a region where before
there was a strong quantum force preventing the approach of
trajectories.
Thus, the central intensity maximum is remarkably enhanced, as can be
seen in Fig.~\ref{fig:3}(b), when compared with the same peak in
Fig.~\ref{fig:3}(a).
This enhancement occurs at the expense of the intensity of the other
maxima.
That is, there is a transfer of trajectories from the outermost to the
innermost diffraction channels.
Dynamically, once the strong boundary imposed by the quantum potential
along the symmetry axis (which is stronger than along the directions
separating the different diffraction channels \cite{Sanz1})
is suppressed, the ``quantum pressure'' \cite{Sanz1,Sanz2} pushes the
trajectories towards the region covered by the empty slit,
and favors their transfer from one diffraction channel to the nearby one.

As $\eta$ is further increased, quantum trajectories are also able
to penetrate more across the symmetry axis, as it is apparent in
Fig.~\ref{fig:3}(c).
The availability of a wider accessible region in the other side of
the symmetry axis makes the concentration of trajectories along
this direction to decrease, and they distribute more homogeneously.
This causes a remarkable decrease of the coherence degree in the
measured intensity, although the central peaks are still relatively
intense.
This trend continues until the interference pattern completely
disappears when the decoupling is maximum [see Fig.~\ref{fig:3}(d)].
In this case, the trajectories are unaffected by the presence of the
other slit, i.e.\ they display a totally classical-like behavior.
Moreover, a full transition from a dynamics characterized by
single-valuedness of the momentum to another where it is bi-valued
is observed, taking place this process within the system subspace.


\section{\label{sec5} Final discussion and conclusions}

Trajectory-based approaches have received much attention in the last
years as a potential tool to handle and study high-dimensional
complex quantum systems \cite{chapter}.
Nowadays the design of new numerical tools highly relies on this
kind of formalisms rather than using other approaches based on the
time-dependent Schr\"odinger equation \cite{wyatt-book}.
In the particular case of Bohmian mechanics, this computational power
combines with its capability to provide causal interpretations to
quantum phenomena within a purely quantum-mechanical framework, unlike
other semiclassical or classical approaches.
These two interesting features have brought Bohmian mechanics from
the field of the Foundations of Physics to the most applied fields
in physics and chemical physics, attracting the interest of many
different communities within.
Similarly, the model proposed here to study decoherence and the process
of going from one context to another one in quantum mechanics provides
an understandable and intuitive insight into dynamics involved.
In particular, regarding the mechanism that leads a linear theory with
single-valued momenta to an ``apparent'' non-linear one, where the
momentum can be multi-valued.

Here we have dealt with the problem of the damping or quenching of the
interference fringes produced by decoherence in a two-slit experiment
under the presence of an environment, which yields as a result a
classical-like pattern.
This effect is, to some extent, similar to consider the transition from
a context where both slits are open simultaneously to another one where
only one slit is open at a time and, therefore, the particle track is
localized (i.e.\ the particle has passed through one slit or the
other).
In order to elucidate and understand decoherence without taking into
account explicitly the dynamics of the environment degrees of freedom,
we have considered some simple reduced quantum-trajectory models.
Though limited because of their simplicity, one can infer from them
that, unless there is a sort of screening of the empty-slit
information, the patterns affected by decoherence can be well
reproduced, but keeping still a sort of internal coherence which
makes the trajectories to avoid their crossing, as in standard
Bohmian mechanics.
Only when the empty-slit information is gradually screened, the
trajectories start to cross, as one would expect when using Bohmian
mechanics and projecting the true $3(N+1)$-dimensional quantum
trajectories onto the subspace of the reduced system.
Thus, though simple and approximate ---more refined and precise models,
which are out of the scope of this work, but that are being currently
developed in order to go beyond the analysis presented here---, this
model allows us to study the properties of decoherence at a relatively
cheap computational cost and providing, at the same time, a physical
insight on the way how the contextual dependence and nonlocal
correlations are gradually suppressed.


\section*{Acknowledgements}

This work has been supported by the Ministerio de Ciencia e Innovaci\'on
(Spain) under Projects MTM2006-15533, CONSOLIDER 2006-32, and
FIS2007-62006; Comunidad de Madrid under Project S-0505/ESP-0158;
and Agencia Espa\~nola de Cooperaci\'on Internacional under Project
A/6072/06.
A.S. Sanz also acknowledges the Consejo Superior de Investigaciones
Cient\'{\i}ficas for a JAE-Doc Contract.


\end{document}